\documentclass[%
prd
 ,twocolumn%
 ,secnumarabic%
,amssymb, amsmath,nobibnotes,showpacs]{revtex4}
\usepackage{bm}%
\expandafter\ifx\csname package@font\endcsname\relax\else
 \expandafter\expandafter
 \expandafter\usepackage
 \expandafter\expandafter
 \expandafter{\csname package@font\endcsname}%
\fi

\begin{document}

\title{Gravity in presence of fermions as a SU(2) gauge theory}%

\author{Francesco Cianfrani$^{1}$, Giovanni Montani$^{123}$}%
\email{montani@icra.it, francesco.cianfrani@icra.it}
\affiliation{$^{1}$ICRA-International Center for Relativistic Astrophysics, Dipartimento di Fisica (G9),\\
Universit\`a  di Roma, ``Sapienza'', Piazzale Aldo Moro 5, 00185 Roma, Italy.\\ 
$^{2}$ENEA C.R. Frascati (Unit\`a Fus. Mag.), Via Enrico Fermi 45, 00044 Frascati, Roma, Italy.\\
$^{3}$ICRANet C. C. Pescara, Piazzale della Repubblica, 10, 65100 Pescara, Italy.}
\date{January 2010}%

\begin{abstract}
The Hamiltonian formulation of the Holst action in the presence of a massless
fermion field with a nonminimal Lagrangian is performed without any
restriction on the local Lorentz frame. It is outlined that the phase-space
structure does not resemble that one of a background-independent
Lorentz gauge theory, as some additional constraints are present.
Proper phase-space coordinates are introduced, such that SU(2)
connections can be defined, and the vanishing of conjugate momenta to boost
variables is predicted. Finally, it is demonstrated that for a particular value of the
nonminimal parameter, the kinematics coincides with that one of a
background-independent SU(2) gauge theory, and the
Immirzi parameter becomes the coupling constant of such an interaction between
fermions and the gravitational field.
\end{abstract}


\maketitle

\section{Introduction} 
The main achievements concerning the formulation of a background-independent quantum theory have been obtained by the loop quantum gravity (LQG) approach \cite{revloop} using techniques proper of lattice gauge theories. This is possible because the Gauss constraints of a SU(2) Yang-Mills gauge theory arise in the Hamiltonian formulation \cite{Ash}. The standard quantization procedure is based on fixing the local Lorentz frame according with the time-gauge condition. Then, a proper representation of the holonomy-flux algebra is developed via the GNS construction \cite{GNS}, and a unique 3-diffeomorphism invariant measure is selected out \cite{ALMMT95}. In this scheme the kinematical Hilbert space can be defined and an impressive result is the prediction of discrete spectra for geometrical operators \cite{discr}.

In a previous work \cite{prl} it has been outlined that the SU(2) gauge structure can be recovered also without any restriction on the local Lorentz frame. In fact, as soon as proper phase-space coordinates are chosen,
boost parameters become nondynamical and the constraints coincide with the SU(2) Gauss ones, together with the vanishing behavior of the super-momentum and of the super-Hamiltonian. The significance of this result overcomes the quantum gravity issue and sheds new light on the nature of gravity as a background-independent gauge theory. Furthermore, such a feature can also offer a new perspective to the interaction with elementary particles. For this reason, the work \cite{prl} has been extended to include a nonminimally coupled scalar field \cite{tgnnm}, an Immirzi field \cite{imm} while here spinor matter is addressed.  The latter case is particularly interesting because spinors are sensitive to boost transformations and they enter the SU(2) Gauss constraints when the time gauge holds \cite{lqgsp,Thmatt}.  

In particular, in this work the Hamiltonian formulation of gravity in presence of a massless spinor field with a nonminimal action is developed (see \cite{mer,alsp} for a second-order analysis of the nonminimal fermion Lagrangian). The set of Hamiltonian constraints is analyzed and its second-class character is recognized. Then, a proper parametrization is given for the constraint hypersurfaces, which generalizes that one discussed for the vacuum case in \cite{prl}. The adopted parametrization allows to infer, via a canonical transformation on initial phase-space coordinates, SU(2) connections and boost variables $\chi_a$, together with their conjugated momenta. By a redefinition of spinors, which formally coincides with a boost transformation, conjugate momenta to $\chi_a$ are constrained to vanish. Hence, boost parameters turn out to be nondynamical also in the presence of a spinor field.

As soon as the super-momentum and the super-Hamiltonian are concerned, we point out the peculiar case in which 
the parameter characterizing the nonminimal spinor Lagrangian is equal to the Immirzi parameter. Hence, the resulting expression for the interaction between the spinor and the geometry coincides with that of a SU(2) gauge theory, a part from the presence of 4-fermion terms into the super-Hamiltonian. Therefore, for a specific expression of the nonminimal action, the kinematics of the fermion-gravity system coincides with that predicted in a background-independent SU(2) gauge theory, whose associated spinor current is the ``boosted'' axial one. 

The manuscript is organized as follows: in Sec. \ref{1}, the Hamiltonian structure is presented, and the set of constraints is recognized as being second class. Then, in Sec. \ref{2}, the reduction to a first-class set takes place by restricting to a specific hypersurfaces of the phase-space. The analysis of constraints on such an hypersurfaces is performed in Sec. \ref{3}, and the emergence of SU(2) Gauss constraints is outlined. Hence, boost parameters turn out to be nondynamical in Sec. \ref{4} by a proper redefinition of the spinor fields, while in Sec. \ref{5} the expressions of the super-Hamiltonian and of the super-momentum in the adopted set of variables are analyzed. Brief concluding remarks follows in Sec. \ref{6}.

In the following the signature adopted will be $(+,-,-,-)$. Capital Latin letters $A, B, C,\ldots$ and lowercase Latin letters $a,b,c,\ldots$ will denote four-dimensional and three-dimensional tangent space indexes, respectively. The latter will be raised by $\eta^{ab}=diag[-1,-1,-1]$ (so $\chi^2=\eta^{ab}\chi_a\chi_b$ will be a negative quantity). The space-time indexes will be $\mu,\nu,\rho,\ldots$ for the full four-dimensional manifold, while we will use $i,j,k,\ldots$ for the three-dimensional spatial hypersurfaces.

\section{Hamiltonian formulation}\label{1}

Let us consider a space-time manifold endowed with a metric tensor $g_{\mu\nu}$. We set the local Lorentz frame by virtue of the 4-bein vectors $e^A_\mu$ and of the spin connections $\omega^{AB}_\mu$. We denote $e^A_\mu$ as follows:
\begin{equation}
e^0_\mu=(N,-\chi_aE^a_i),\quad e^a_\mu=(N^iE_i^a,E^a_i).
\end{equation}
Here, $\chi_a$ determine the velocity components of the local Lorentz frame with respect to the spatial manifold in a 3+1 representation. We will work with arbitrary $\chi_a$ space-time functions, such that no restriction on the Lorentz frame holds. 

A massless spinor field is introduced, and the associated Lagrangian density is developed by extending the Dirac formalism to a curved space-time \cite{bw}. This procedure implies the replacement of partial derivatives with covariant ones, {i.e.}
\begin{equation}
D_\mu\psi=\partial_\mu\psi-\frac{i}{2}\omega^{AB}_{\phantom{12}\mu}\Sigma_{AB}\psi,\qquad \Sigma_{AB}=\frac{i}{4}[\gamma_A,\gamma_B],
\end{equation}
$\Sigma_{AB}$ being Lorentz group generators.

We describe the gravitational sector by virtue of the Holst Lagrangian formulation\cite{Ho96}. We demonstrated in \cite{prl} that such a picture allows us to infer the phase-space structure proper of a $SU(2)$ gauge theory in vacuum. The resulting dynamical system can be described by the following action (in units $8\pi G=\hbar c=1$): 
\begin{eqnarray}  
S=\int \sqrt{-g}\bigg[e^\mu_A e^\nu_B R_{\mu\nu}^{AB}-\frac{1}{2\gamma}e^\mu_A e^\nu_B\epsilon^{AB}_{\phantom1\phantom2CD}R_{\mu\nu}^{CD}+\nonumber\\+\frac{i}2(\bar{\psi}\gamma^\mu AD_\mu\psi-D_\mu\bar{\psi}A\gamma^\mu\psi)\bigg]d^4x,\label{act}
\end{eqnarray}
where the expression for $R^{AB}_{\mu\nu}$ reads
\begin{equation}
R^{AB}_{\mu\nu}=\partial_{[\mu}\omega^{AB}_{\nu]}-\omega^A_{\phantom1C[\mu}\omega^{CB}_{\nu]},
\end{equation}
while the matrix $A$ contains the parameter $\alpha$ giving the nonminimal term as follows:
\begin{equation}
A=1+i\alpha\gamma_5.
\end{equation} 
Here, $\gamma$ denotes the Immirzi parameter, which is a fundamental quantum ambiguity arising in vacuum LQG \cite{RT98}. In the presence of spinors, such an ambiguity is promoted to be a classical feature, since it determines the amount of the 4-fermions interaction term arising after the II Cartan structure equations have been solved \cite{PR06}.

We take $\omega^{AB}_{\phantom{12}\mu}$, $\psi$ and $\bar\psi$ as configuration variables and conjugate momenta are given by the following expressions:
\begin{eqnarray}
{}^\gamma\!\pi^t_{AB}=0,\\ {}^\gamma\!\pi^i_{AB}=\pi^i_{AB}-\frac{1}{2\gamma}\epsilon^{CD}_{\phantom{12}AB}\pi^i_{CD},\qquad(\pi^i_{AB}=\sqrt{-g}e^t_{[A}e^i_{B]}),\\ \bar\Pi=\frac{i}{2}\sqrt{-g}\bar\psi\gamma^AAe_A^t,\qquad \Pi=-\frac{i}{2}\sqrt{-g}e_A^tA\gamma^A\psi,\label{consp}
\end{eqnarray}
respectively. Therefore, $\omega^{AB}_{\phantom{12}t}$ are not dynamical variables, and in the following we will treat them as Lagrangian multipliers. Furthermore, the conditions (\ref{consp}) do not commute among each other, and they can be solved by writing explicitly $\Pi$ and $\bar{\Pi}$ in terms of $\psi$ and $\bar\psi$, at the same time using the following commutation relation: 
\begin{equation}
[\psi,\bar\psi]=-\frac{i}{\sqrt{h}\sqrt{1+\chi^2}}(\gamma^0+\chi_a\gamma^a).\label{psicom}
\end{equation}
There are six further conditions coming from the definitions of conjugate momenta, {\it i.e.}
\begin{equation}
C^{ij}=\epsilon^{ABCD}\pi^{(i}_{AB}\pi^{j)}_{CD}=0.\label{C}
\end{equation}
As soon as a Legendre transformation is performed and the Lagrangian multipliers $\lambda^{AB}$, $\lambda_{ij}$ and $\eta_{ij}$, are introduced, the Hamiltonian in a 3+1 representation for the metric tensor can be written as 
\begin{equation}   \mathcal{H}=\int\bigg[\widetilde{N}H+\widetilde{N}^iH_i+\lambda^{AB}G_{AB}+\lambda_{ij}C^{ij}+\eta_{ij}D^{ij}\bigg]d^3x,\end{equation}
$\widetilde{N}$ and $\widetilde{N}^i$ being the lapse function and the shift vector, respectively. Since $\widetilde{N}$ and $\widetilde{N}^i$ are nondynamical, the expression above turns out to be a linear combination of constraints and the full Hamiltonian vanishes. In particular, the constraints are
\begin{itemize}
{\item the super-Hamiltonian one, whose expression is
\begin{eqnarray}   
H=\frac1{\sqrt{h}}\left(\delta^{CD}_{AB}-\frac{1}{2\gamma}\epsilon^{CD}_{\phantom{12}AB}\right)\pi^i_{CF}\pi^{jF}_{\phantom1D}R^{AB}_{ij}+\nonumber\\+\frac{i}{\sqrt{h}} \pi^i_{AB}(\bar\Pi\Sigma^{AB}D_i\psi-D_i\bar\psi\Sigma^{AB}\Pi)=0, \label{sh}
\end{eqnarray}
$h_{ij}$ and $h$ being the spatial metric tensor and the associated determinant, respectively; 
}
{\item the super-momentum ones, {\it i.e.}
\begin{eqnarray}
H_i=\left(\delta^{CD}_{AB}-\frac{1}{2\gamma}\epsilon^{CD}_{\phantom{12}AB}\right)\pi^j_{CD}R^{AB}_{ij}+\nonumber\\+\bar\Pi D_i\psi+D_i\bar\psi\Pi=0;\label{sm}
\end{eqnarray}
}
{\item the Gauss constraints of the local Lorentz group, which take the following form:
\begin{eqnarray}
G_{AB}=\partial_i{}^\gamma\!\pi^i_{AB}-2\omega_{[A\phantom2i}^{\phantom1C}{}^\gamma\!\pi^i_{|C|B]}-\nonumber\\-i(\bar\Pi\Sigma_{AB}\psi-\bar\psi\Sigma_{AB}\Pi)=0;
\end{eqnarray}
}
{\item $C^{ij}=0$ and the associated secondary ones 
\begin{eqnarray}
D^{ij}=\epsilon^{ABCD}\pi^k_{AF}\pi^{(i|F|}_{\phantom1\phantom2B}D_k\pi^{j)}_{CD}+\frac{\gamma^2}{2(\gamma^2+1)}\bigg(\epsilon^{AB}_{\phantom{12}CD}-\nonumber\\-\frac{2}{\gamma}\delta^{AB}_{CD}\bigg)\pi^{(i}_{AB}\pi^{j)}_{FG}(\bar\Pi\Sigma^{FG}\Sigma^{CD}\psi+\bar\psi\Sigma^{CD}\Sigma^{FG}\Pi)=0.
\label{D}
\end{eqnarray}
}
\end{itemize}
Hence, the super-Hamiltonian and the super-momentum vanish and this feature is due to the invariance of the action (\ref{act}) under 4-diffeomorphisms. Furthermore, the presence of the Gauss constraints associated to the Lorentz group makes this formulation close to the analogous one for a Yang-Mills gauge theory. However, the additional conditions (\ref{C}) and (\ref{D}) make the whole set of constraints second class, which means that some variables are redundant, and a nontrivial symplectic structure is induced on the constraint hypersurfaces. We will see in the next sections that as soon as a proper set of variables is adopted, by which we can avoid (\ref{C}) and (\ref{D}), the Gauss constraints of a SU(2) gauge theory will be obtained.    
  
\section{Analysis of $C^{ij}=0$ and $D^{ij}=0$}\label{2}
The hypersurfaces defined by conditions (\ref{C}) and (\ref{D}) can be parametrized by fixing $\omega^{ab}_{\phantom{12}i}$ and $\pi^i_{ab}$. In the following we will denote $\pi^i_{0a}$ by $\pi^i_a$ and introduce their inverses $\pi^b_j$. The 3-metric of the spatial manifold is given by the following expression 
\begin{equation}
h_{ij}=-\frac{1}{\pi}T^{-1}_{ab}\pi^a_i\pi^b_j, \qquad T^{-1}_{ab}=\eta_{ab}+\chi_a\chi_b,
\end{equation}
$\pi$ being the determinant of $\pi^a_i$. We can associate to $\pi^i_a$ the quantity ${}^\pi\!\omega^{\phantom1b}_{a\phantom1i}$, defined as follows
\begin{equation}
{}^\pi\!\omega^{\phantom1b}_{a\phantom1i}=\frac{1}{\pi^{1/2}}\pi^b_l{}^3\!\nabla_i(\pi^{1/2}\pi^l_a),
\end{equation}
${}^3\!\nabla_i$ being the covariant derivatives associated to $h_{ij}$.

A proper solution to constraints (\ref{C}) and (\ref{D}) is the following one
\begin{eqnarray}
\pi^i_{ab}=2\chi_{[a}\pi^i_{b]},\label{scon1}\\ \omega^{\phantom1b}_{a\phantom1i}={}^\pi\!\omega^{\phantom1c}_{a\phantom1i}T^{-1b}_c+\chi_a\omega^{0b}_{\phantom{12}i}+\chi^b
(\omega_{a\phantom1i}^{\phantom10}-\partial_i\chi_a)+{}^\psi\!\omega^{\phantom1b}_{a\phantom1i},\label{scon2}
\end{eqnarray}
where ${}^\psi\!\omega^{\phantom1b}_{a\phantom1i}$ is the modification with respect to the vacuum case and it reads
\begin{eqnarray}
{}^\psi\!\omega^{ab}_{\phantom1i}=+\frac{1}{4}\frac{\gamma(\gamma-\alpha)}{\gamma^2+1\sqrt{i+\chi^2}}\epsilon^{ab}_{\phantom{12}c}\pi^c_i (J^0+\chi_d J^d)-\nonumber\\-\frac{1}{2}\frac{\gamma(1+\alpha\gamma)}{\gamma^2+1}\pi^c_i T^{-1[a}_c\eta^{b]d}(J_d-\chi_dJ^0),
\end{eqnarray}
$J^A$ being $J^A=\sqrt{h}\bar{\psi}\gamma_5\gamma^A\psi$.

By virtue of such a modification the boost constraints are not independent conditions anymore, since by substituting the expressions (\ref{scon1}) and (\ref{scon2}) into $G_{AB}$ we find
\begin{equation}
G_{0a}=\chi^bG_{ab}.
\end{equation}
The possibility to avoid the boost constraints, once the rotational ones hold, will allow us to define a set of canonical coordinates on the hypersurfaces $C^{ij}=D^{ij}=0$. At the same time, $\chi_a$ will be promoted to be configuration variables, such that the arbitrariness of the local Lorentz frame is preserved.

\section{Analysis of the Gauss constraints of the Lorentz group}\label{3}
As soon as we replace $\pi^i_{ab}$ and $\omega^{ab}_{\phantom1i}$ by the expressions (\ref{scon1}) and (\ref{scon2}),  we remove the constraints (\ref{C}) and (\ref{D}) from the set of phase-space constraints, and the boost ones become redundant conditions. However, because the original constraints are second class, a nontrivial symplectic structure is induced, and it is the starting point to select out a proper set of phase-space variables.  
  
The first couple of conjugate variables is inferred by setting densitized 3-bein vectors of the spatial metric, $\widetilde{\pi}^i_a$, as  momenta, as in the standard formulation of LQG \cite{revloop}. In particular the expressions of $\widetilde{\pi}^i_a$ in terms of $\pi^i_a$ read as follows:
\begin{equation}
\widetilde{\pi}^i_a=S^b_a\pi^i_b,\qquad S^a_b=\sqrt{1+\chi^2}\delta^a_b+\frac{1-\sqrt{1+\chi^2}}{\chi^2}\chi_a\chi_b.
\end{equation}
The associated canonical conjugate variables are given by 
\begin{eqnarray}
\widetilde{A}_i^a=S^{-1a}_b\bigg((1+\chi^2)T^{bc}\left(\omega_{0ci}+{}^\pi\!D_i\chi_c\right)-\nonumber\\-\frac{1}{2\gamma}\epsilon^b_{\phantom1cd}({}^\pi\!\omega^{cf}_{\phantom1\phantom2i}
T^{-1d}_{\phantom1f}+{}^\psi\!\omega^{cd}_{\phantom{12}i})+\nonumber\\+\frac{2+\chi^2-2\sqrt{1+\chi^2}}{2\gamma\chi^2}\epsilon^{bcd}\partial_i\chi_c\chi_d\bigg),\quad\label{ABI}
\end{eqnarray}
where ${}^\pi\!D_i\chi_c=\partial_i\chi_c-{}^\pi\!\omega_{c\phantom1i}^{\phantom1b}\chi_b-\frac{1}{1+\chi^2}{}^\psi\!\omega_{c\phantom1i}^{\phantom1b}\chi_b$. 

The expressions (\ref{ABI}) are the generalization of Ashtekar-Barbero-Immirzi variables \cite{Ash} to a generic local Lorentz frame in presence of spinor fields. It is worth noting how the spinor enters the definitions of $\widetilde{A}_i^a$, even though the latter is a quantity characterizing the geometry. This feature is a consequence of the significant modification such a matter field provides to the space-time structure: the introduction of torsion.

Other phase-space variables are $\chi_a$ themselves and the conjugate momenta $\pi^a$.   

The constraints $G_{ab}=0$ can now be written in the new set of coordinates, and they can be shown to be equivalent to the following expression
\begin{equation}
\partial_i\widetilde{\pi}^i_a+\gamma\epsilon_{ab}^{\phantom{12}c}\widetilde{A}^b_i\widetilde{\pi}_c^i=-\frac{\gamma}{2\sqrt{1+\chi^2}}S_a^b(J_{b}-\chi_b J_0).\label{G1}
\end{equation}
It is impressive that the conditions above reduce to the usual Gauss constraints of a SU(2) gauge theory, when the spinor fields are absent. However, the source term is made up of two commuting variables, boost parameters and spinors.  

\section{The ``boosted'' spinor and the nonevolutionary character of $\chi^a$}\label{4}
Other conditions are inferred by imposing that the transformation $\{\omega^{AB}_{\phantom{12}i},{}^\gamma\!\pi^i_{AB},\psi,\bar{\psi}\}\rightarrow\{\widetilde{A}^a_i,\widetilde{\pi}^i_a,\chi_a,\pi^a,\psi,\bar{\psi}\}$ is canonical, {\it i.e.}
\begin{eqnarray}
\frac{1}{2}{}^\gamma\!\pi^i_{AB}\partial_t\omega^{AB}_{\phantom{12}i}+\bar\Pi\partial_t\psi+\partial_t\bar\psi\Pi=\nonumber\\=\widetilde{\pi}^i_a\partial_t\widetilde{A}^a_i+\pi^a\partial_t\chi_a+\bar\Pi\partial_t\psi+\partial_t\bar\psi\Pi.\label{sf}
\end{eqnarray}
Such a requirement fixes $\pi^a$ as follows 
\begin{equation}
\pi^a=-i\frac{1-\sqrt{1+\chi^2}}{\chi^2}\chi^b(\bar\Pi\Sigma_{a}^{\phantom1b}\psi-\bar\psi\Sigma_{a}^{\phantom1b}\Pi).
\end{equation} 
Unlike the vacuum case \cite{prl}, $\pi^a$ do not vanish. Hence the variables $\chi^a$ are dynamical. However, this feature just reflects the fact that spinor fields are sensitive to $\chi_a$ changes. In fact, a nondynamical term can be defined and it involves a combination of $\pi^a$ and spinor variables. This can be seen by performing the following redefinition:
\begin{equation}
\psi=e^{i\chi^a\Sigma_{0a}}\psi^*.\label{boo}
\end{equation}
By writing $\bar\Pi\partial_t\psi+\partial_t\bar\psi\Pi$ in terms of $\psi^*$ and $\bar\psi^*$, a contribution $i\frac{1-\sqrt{1+\chi^2}}{\chi^2}\chi^b(\bar\Pi\Sigma_{a}^{\phantom1b}\psi-\bar\psi\Sigma_{a}^{\phantom1b}\Pi)\partial_t\chi_a$ is added to the full symplectic form (\ref{sf}), such that the new conjugate momenta to $\chi_a$ are constrained to vanish. Hence, the choice of $\psi^*$ and $\bar\psi^*$ as phase-space variables makes $\chi_a$ nondynamical.

The redefinition (\ref{boo}) acts on spinor variables as a boost to the Lorentz frame where the time gauge holds. In fact, any spinor quantity in terms of $\psi^*$ looks like the quantity evaluated in the time gauge, such that there is no dependence on $\chi_a$. For instance, the commutation relations between $\psi^*$ and $\bar\psi^*$ read
\begin{equation} 
[\psi^*,\bar\psi^*]=-\frac{i}{\sqrt{h}}\gamma^0,
\end{equation}
which coincides with the expressions that the Poisson brackets (\ref{psicom}) take in the time gauge. Similarly, the source term into the constraints (\ref{G1}) becomes the SU(2) currents one finds for $\chi_a=0$. 

Hence, as soon as the spinor fields are written in terms of the proper components $\psi^*$, living in the time-gauged Lorentz frame, the Gauss constraints of the local Lorentz group and the conditions $(\ref{C})$ and $(\ref{D})$ reduce to the following system:
\begin{equation}
\partial_i\widetilde{\pi}^i_a+\gamma\epsilon_{ab}^{\phantom{12}c}\widetilde{A}^b_i\widetilde{\pi}_c^i=-\frac{\gamma}{2}J^*_{a},\qquad \pi^a=0,\label{Gc}
\end{equation}
where $J^*_{A}=\sqrt{h}\bar{\psi}^*\gamma_A\gamma_5\psi^*$ is the axial component of the fermion current.

Therefore, without fixing the local Lorentz frame and for any value of $\alpha$, \emph{the Gauss constraints of a SU(2) gauge theory are inferred, whose source is the boosted axial current, while boost parameters are nondynamical}.

\section{The super-momentum and the super-Hamiltonian constraints}\label{5}
In the adopted set of variables the super-momentum constraints take the following expressions
\begin{eqnarray}
H_i=\widetilde{\pi}^j_aF^a_{ij}+\frac{i}{2}\sqrt{h}(\bar{\psi}^*\gamma^0A\partial_i\psi^*-\partial_i\bar\psi^*A\gamma^0\psi^*)+\nonumber\\+\frac{\alpha}{2}\widetilde{A}^a_iJ^*_{a}+\frac{\alpha-\gamma}{4\gamma}\epsilon^a_{\phantom1bc}\widetilde{\omega}^{bc}_iJ^*_{a}-\frac{(\alpha-\gamma)^2}{4(\gamma^2+1)}J^*_0J^*_{a}\widetilde{\pi}^a_i,\label{ssm}
\end{eqnarray}
where $F^a_{ij}=\partial_{[i}\widetilde{A}^a_{j]}+\gamma\epsilon^a_{\phantom1bc}\widetilde{A}^b_i\widetilde{A}^c_j$ is the field strength of the SU(2) connection $\widetilde{A}^a_i$, while $\widetilde{\omega}^{ab}_i$ are spin connections associated to $\widetilde{\pi}^i_a$, {\it i.e.}
\begin{equation}
\widetilde{\omega}^{ab}_i=\widetilde{\pi}_j^b{}^3\!\nabla_i\widetilde\pi^j_c\eta^{ac}.
\end{equation}
By avoiding a term proportional to the Gauss constraints (\ref{Gc}), the super-Hamiltonian can be written as 
\begin{eqnarray}
H=\frac{\widetilde{\pi}^i_a\widetilde{\pi}^j_b}{2\sqrt{h}}\epsilon^{\phantom{12}c}_{ab}\widetilde{F}^c_{ij}-\frac{(\gamma^2+1)}{\gamma^2\sqrt{h}}\widetilde{\pi}^i_a\widetilde{\pi}^j_b
\big(\partial_{[i}\widetilde\omega^{ab}_{\phantom{12}j]}-\widetilde\omega^{ac}_{\phantom{12}[i}\widetilde\omega^{\phantom1b}_{c\phantom1j]}\big)-\nonumber\\-\frac{i}{2}\widetilde{\pi}^i_a\left(\bar{\psi}^*\gamma^aA\partial_i\psi^*-\partial_i\bar\psi^*A\gamma^a\psi^*\right)+\nonumber\\+\frac{\alpha}{2\sqrt{h}}\widetilde\pi^i_a\widetilde{A}^a_iJ^*_0-\frac{\alpha\gamma}{2\sqrt{h}}\widetilde\pi^i_a\epsilon^{ab}_{\phantom{1b}c}\widetilde{A}^c_iJ^*_b+\frac{\alpha-\gamma}{4\gamma\sqrt{h}}\epsilon^a_{\phantom1bc}\widetilde{\pi}^i_a\widetilde{\omega}^{bc}_iJ^*_0-\nonumber\\-\frac{3}{16\sqrt{h}}\frac{(\gamma-\alpha)^2}{\gamma^2+1}(J^*_0)^2-\frac{(1+\alpha\gamma)(4\gamma^2-3\alpha\gamma+1)}{16(\gamma^2+1)\sqrt{h}}J^*_cJ^{*c}\label{ssh}.
\end{eqnarray}
It is worth noting that the interaction between the spinor fields and $\widetilde{A}^a_i$ is exactly the one for a SU(2) gauge theory having the axial current as the source. However, in general the similarity with a Yang-Mills model is weakened by the presence of terms containing $\widetilde{\omega}^{ab}_i$, so momenta $\widetilde{\pi}^i_a$. 

For instance these terms are present in two relevant cases: the minimal one ($\alpha=0$) and the nonminimal one, which reproduces the Einstein-Cartan theory ($\alpha=1/\gamma$) \cite{mer}.

A significant simplification for $H_i$ and $H$ occurs by fixing the nonminimal parameter equal to the Immirzi parameter, {\it i.e.}
\begin{equation}
\alpha=\gamma.
\end{equation}
In fact, in this case one finds a ``pure'' SU(2) interaction between the geometry and the spinor, since the expressions (\ref{ssm}), (\ref{ssh}) can be rewritten as
\begin{eqnarray}
H_i=H^G_i+\frac{i}{2}\sqrt{h}(\bar{\psi}^*\gamma^0A{}^{(A)}\!D_i\psi^*-{}^{(A)}\!D_i\bar\psi^*A\gamma^0\psi^*),\\
H=H^G-\frac{i}{2}\widetilde{\pi}^i_a\bigg(\bar{\psi}^*\gamma^aA{}^{(A)}\!D_i\psi^*-{}^{(A)}\!D_i\bar\psi^*A\gamma^a\psi^*\bigg)-\nonumber\\-\frac{1+\gamma^2}{16\sqrt{h}}{}^A\!J_c{}^A\!J^c.
\end{eqnarray}
where ${}^{(A)}\!D_i\psi=\partial_i\psi-\frac{i}{2}\gamma\widetilde{A}^a_iT_a\psi$ and the gauge generator is $T_a=\epsilon_a^{\phantom{12}bc}\Sigma_{bc}$. The Immirzi parameter finds a natural interpretation as the coupling constant of this SU(2) interaction.

The only differences with a SU(2) gauge theory \`a-la Yang-Mills are given by\begin{itemize}{\item the free part of the gravitational super-Hamiltonian $H^G$;}{\item the presence of the 4-fermion terms, which makes the theory perturbative nonrenormalizable.} \end{itemize} 

Therefore, the case $\gamma=\alpha$ can be regarded as outstanding because the kinematics of the gravity-fermion system is that of a background-independent SU(2) gauge theory and for the interpretation that $\gamma$ finds in this scheme.
 
\section{Conclusions}\label{6}
We analyzed the phase-space of gravity in the presence of a massless spinor field described by a nonminimal action in a generic local Lorentz frame. We pointed out that Hamiltonian constraints belong to a second-class set. In order to identify physical degrees of freedom, a proper parametrization was chosen, which enabled us to solve some constraints and to select SU(2) connections and boost parameters as configuration variables. By means of a proper redefinition of the spinor fields, it was shown that conjugate momenta to boost parameters vanish, so $\chi_a$ did not have an evolutionary character. Hence, the final set of constraints was formed by the Gauss conditions of the SU(2) group, the super-momentum and the super-Hamiltonian constraints. This result allows to apply the LQG quantization procedure even in presence of spinors, independently of whether or not we fix the local Lorentz frame, along the lines of \cite{lqgsp,Thmatt}.

We also demonstrated that fixing the nonminimal parameter $\alpha$ equal to the Immirzi one, the interaction between gravity and the spinor fields becomes that of a background-independent SU(2) gauge theory and the associated gauge transformations are generated by rotations. This feature makes the whole Hamiltonian formulation closer to the one of other fundamental interactions than could be expected from the Lagrangian symmetries only. Furthermore, the Immirzi parameter can be viewed as the coupling constant of such an interaction. 

In view of the emergence of 4-fermions interaction terms, which provide us with perturbative nonrenormalizable contributions, a natural framework in which analyzing the proposed scenario is a path-integral formulation. In this respect, the reduction we performed of the full local Lorentz invariance to the SU(2) one is promising for the definition of a well-grounded measure and the interpretation of the issues of our work in the context of spin-foam models \cite{SF} is an exciting perspective.


\begin{thebibliography}{99}

\bibitem{revloop}
C. Rovelli, ``Quantum gravity'', Cambridge University Press, Cambridge, (2004), XXIII; T. Thiemann, ``Modern Canonical Quantum General Relativity'', (Cambridge University Press, Cambridge, England, 2006); F. Cianfrani, O.M. Lecian, G. Montani, ``Fundamentals and recent developments in non-perturbative canonical Quantum Gravity'', arXiv:0805.2503.

\bibitem{Ash}
A. Ashtekar, {\it Phys. Rev. Lett.}, {\bf 57}, (1986), 2244; {\it Phys. Rev. D}, {\bf 36}, (1987), 1587;
J. F. Barbero, {\it Phys. Rev. D}, {\bf 51}, (1995), 5507. 

\bibitem{GNS}
R. Haag, ``Local quantum physics : fields, particles, algebras'' - Berlin : Springer-Verlag, (1992) - XIV.

\bibitem{ALMMT95}
A. Ashtekar and J. Lewandowski, {\it J. Geom. Phys.}, {\bf 17}, (1995), 191-230; 
J. Lewandowski, A. Okolow, H. Sahlmann, T. Thiemann, {\it Comm. Math. Phys.}, {\bf 267}, No. 3,(2006) 703-733. 

\bibitem{discr}
C. Rovelli, L. Smolin, {\it Nucl. Phys. B}, {\bf 442}, (1995), 593-622; Erratum-ibid. {\bf 456}, (1995), 753;
A. Ashtekar, J. Lewandowski, {\it Class. Quant. Grav.}, {\bf 14}, (1997), A55-A82.



\bibitem{prl}
F. Cianfrani, G. Montani, {\it Phis. Rev. Lett.}, {\bf 102}, (2009), 091301; `` The Role of Time-Gauge in Quantizing Gravity'', to appear on proceedings of the III Stueckelberg workshop, Pescara July 8-18 (2008), arXiv:0904.0573; 

\bibitem{tgnnm}
F. Cianfrani, G. Montani, {\it Phis. Rev. D}, {\bf 80}, (2009), 084045.

\bibitem{imm}
F. Cianfrani, G. Montani, {\it Phis. Rev. D}, {\bf 80}, (2009), 084040.

\bibitem{lqgsp}
M. Bojowald, R. Das, {\it Phys. Rev. D}, {\bf 78}, (2008), 064009.

\bibitem{Thmatt}
T. Thiemann, {\it Class. Quant. Grav.}, {\bf 15}, (1998), 1281-1314; {\it Class. Quant. Grav.}, {\bf 15}, (1998), 1487-1512.

\bibitem{mer}
S. Mercuri, {\it Phys. Rev. D}, {\bf 73}, (2006), 084016. 

\bibitem{alsp}
S. Alexandrov, {\it Class. Quant. Grav.}, {\bf 25}, (2008), 145012. 

\bibitem{bw}
D. R. Brill, J. A. Wheeler, {\it Rev. Mod. Phys.}, {\bf 29}, (1957), 465-479.

\bibitem{Ho96}
S. Holst, {\it Phys. Rev. D}, {\bf 53}, (1996) 5966-5969. 

\bibitem{RT98}
C. Rovelli, T. Thiemann, {\it Phys. Rev. D}, {\bf 57}, (1998), 1009-1014.

\bibitem{PR06}
A. Perez, C. Rovelli, {\it Phys. Rev. D}, {\bf 73}, (2006), 044013. 

\bibitem{SF} J. Engle, R. Pereira, C. Rovelli, {\it Phys. Rev. Lett.}, {\bf 99}, (2007), 161301; J. Engle, E. Livine, R. Pereira, C. Rovelli, {\it Nucl. Phys. B}, {\bf 799}, (2008), 136-149. 

\end{thebibliography}
\end{document}